\documentclass[aps,prb,twocolumn,a4]{revtex4-1}
\usepackage{graphicx}
\usepackage{amsmath,amsfonts,amssymb}
\usepackage[colorlinks,citecolor=blue]{hyperref}
\usepackage{tikz}
\usetikzlibrary{positioning}

\newcommand{\dd}{\mathrm{d}}

\newcommand{\smo}{SrMnO$_3$~}

\begin{document}

\title{Magnetic exchange interactions in SrMnO$_3$}

\author{Xiangzhou Zhu}
\affiliation{Materials Theory, ETH Z\"urich, Wolfgang-Pauli-Strasse 27, 8093 Z\"urich, Switzerland}
\author{Alexander Edstr\"om}
\affiliation{Materials Theory, ETH Z\"urich, Wolfgang-Pauli-Strasse 27, 8093 Z\"urich, Switzerland}
\author{Claude Ederer}
\email{claude.ederer@mat.ethz.ch}
\affiliation{Materials Theory, ETH Z\"urich, Wolfgang-Pauli-Strasse 27, 8093 Z\"urich, Switzerland}

\date{\today}

\begin{abstract}
We calculate Heisenberg-type magnetic exchange interactions for SrMnO$_3$ under isotropic volume expansion using an approach that is based on total energy variations due to infinitesimal spin rotations around a given reference state.
Our total energy calculations using density functional theory (DFT) indicate a transition from antiferromagnetic to ferromagnetic coupling for increasing interatomic distances, corresponding to a sign change of the nearest neighbor exchange interaction. 
This sign change cannot easily be understood from a standard superexchange mechanism. Furthermore, the exchange interaction strongly depends on the corresponding reference state.
This ``non-Heisenberg'' behavior increases with increasing volume and is also confirmed through non-collinear DFT calculations. An orbital- and energy-resolved decomposition of the exchange coupling suggests that an increased partial occupancy of $e_{g}$ orbitals near the Fermi level is crucial both for the sign change and the non-Heisenberg behavior of the nearest neighbor interaction. 
Furthermore, even though both $e_g$ and $t_{2g}$ contributions to the exchange interactions decay exponentially for large inter-atomic distances, the $e_g$ contribution remains surprisingly strong over relatively large distances along the crystal axes.
\end{abstract}

\maketitle

\section{Introduction}

Perovskite structure \smo has received great interest after Lee and Rabe computationally revealed the possibility of strain-engineered multiferroicity in this otherwise G-type antiferromagnetic (AFM), paraelectric compound~\cite{Lee/Rabe:2010}. They predicted that biaxial tensile strain leads to ferroelectricity at a critical strain of around 1\%. They also predicted that strain will induce a series of magnetic transitions, possibly leading to a rare ferromagnetic (FM) ferroelectric (FE) phase at low temperatures around 4\,\% strain. The strain-induced ferroelectricity has later been experimentally corroborated~\cite{Becher2015,Guzman_et_al:2016}, with recent work showing the possibility of growing highly strained films on which FE hysteresis loops have been measured~\cite{Guo2018}.
Experimental evidence for a strain-induced magnetic transition has also been reported~\cite{Maurel_et_al:2015}.
More recent computational work investigated the strain-temperature phase diagram of \smo~\cite{Edstrom/Ederer:2018} and predicted the existence of a tetracritical point, where the magnetic and FE critical temperatures coincide, allowing for intriguing magnetoelectric coupling phenomena~\cite{2019arXiv190512955E}.

As shown in Ref.~\onlinecite{Edstrom/Ederer:2018}, the magnetic transitions under strain can be understood in terms of the in-plane and out-of-plane nearest neighbor magnetic exchange interactions changing sign from AFM to FM coupling for increasing tensile epitaxial strain and  as a result of the ferroelectric displacements, respectively.
The latter can in principle be explained by the distortion of the out-of-plane Mn---O---Mn bond angle, which deviates from 180$^\circ$ due to off-centering of the Mn cation.
In contrast, the sign change of the in-plane coupling under tensile strain cannot easily be understood by simple superexchange arguments.
In this case, the Mn---O---Mn bond angle remains at the ideal value of 180$^\circ$ and only the Mn---O bond lengths change. 

As we show in this work, a similar sign change also occurs in cubic \smo under isotropic volume expansion, whereby ferromagnetism, surprisingly, becomes favored over antiferromagnetism above a certain critical volume.
Furthermore, even though \smo is a magnetic insulator, expected to be relatively well described by the Heisenberg Hamiltonian, with exchange parameters independent of the magnetic reference state, a notable configuration dependence of the exchange interactions has been observed in this material~\cite{Edstrom/Ederer:2018}. 

To analyze the origin of these unexpected features of the magnetic coupling in SrMnO$_3$, we use an approach where magnetic exchange couplings are computed via small deviations from a suitably chosen reference configuration, usually the magnetic ground state~\cite{Liechtenstein:1987,Rudenko:2013,korotin:2015,logemann:2017}. This method allows to analyze orbital- and energy-resolved contributions to the magnetic coupling constants for different reference states, and also to take into consideration further neighbor interactions without the need for large supercells. 

We find that both the non-Heisenberg behavior and the sign change of the nearest neighbor interaction to FM coupling are related to Mn $e_g$ states located below the gap/Fermi level. These Mn $e_g$ states are partially occupied, due to strong hybridization with the O $p$ states, in spite of the formal Mn$^{4+}$ configuration. Volume expansion lowers the energy of these $e_g$ levels and strongly enhances their contribution to the exchange coupling in the energy range immediately below the Fermi level. This causes the change from AFM to FM coupling and increases the deviations from ideal Heisenberg behavior. 
Our results provide important insights into the mechanism underlying the magnetic exchange interactions in SrMnO$_3$. We also point out that similar effects can be observed in a number of other materials.

In the following, we first describe the computational method we use to obtain the magnetic exchange couplings in Sec.~\ref{sec:comp.meth}. Then, we start the presentation of our results obtained for \smo by demonstrating the change from an AFM to FM magnetic ground state under volume expansion, and comparing the nearest neighbor exchange interactions calculated using different methods and reference states in Sec.~\ref{sec:J1vol}. The orbital- and energy-relsolved contributions to the nearest neighbor interaction are discussed in Sec.~\ref{sec:orbres}, and results for further neighbor exchange interactions are presented in Sec.~\ref{sec:furtherneigh}. Finally our main conclusions are summarized in Sec.~\ref{sec:concl}.

\section{Computational Method}\label{sec:comp.meth}

\subsection{Calculation of exchange couplings}

Magnetic exchange interactions can be defined by mapping total energies obtained from first principles electronic structure calculations on a classical Heisenberg Hamiltonian:
\begin{equation}
    H_\text{mag} = - \sum_{i<j} J_{ij} \hat{m}_{i} \cdot \hat{m}_{j} \quad .
\end{equation}
Here, we use normalized unit vectors $\hat{m}_i$ to indicate the direction of the magnetic moment (spin) at site $i$, the summation is over all pairs $i$ and $j$, and the sign convention is such that a positive $J_{ij}$ corresponds to ferromagnetic coupling (i.e., lower energy for parallel spins).

In the following we use two different methods to obtain the coupling constants $J_{ij}$. The first is based on calculating total energy differences of different configurations where either spin $i$ or $j$ (or both) are flipped relative to a chosen reference configuration~\cite{Xiang2011}. 
The exchange couplings are then obtained from:
\begin{equation}
\label{eq:Jij-energy}
    J_{ij} = \frac{E_{\uparrow\downarrow}+E_{\downarrow\uparrow}-E_{\uparrow\uparrow}-E_{\downarrow\downarrow}}{4n} \quad ,
\end{equation}
where the arrows indicate the spin directions of site $i$ and $j$ in the configuration with energy $E_{\sigma_i\sigma_j}$, and $n$ is the number of equivalent bonds between sites $i$ and $j$ within the supercell. 
In practice, this method requires to use a sufficiently large supercell, such that interactions between spin $i$ and the periodic replicas of spin $j$ are negligible.

The second method we use for calculating exchange interactions goes back to the work of Liechtenstein \emph{et al.}~\cite{Liechtenstein:1987}, and corresponds to energy variations due to infinitesimal local spin rotations, also defined relative to a fixed reference configuration.
This reference configuration typically is the magnetic ground state, but can in principle be any stationary state with respect to variations of the spin density, which allows application of the magnetic force theorem.
In this work, only such stationary (and collinear) magnetic configurations are considered.
Within a localized, tight-binding-like, orbital basis set, the exchange interaction between the local magnetic moments on sites $i$ and $j$ can then be expressed as:
\begin{widetext}
\begin{equation}\label{eq:J_ij}
    J_{ij} = \pm \frac{1}{2\pi}\mathrm{Im}\int_{-\infty}^{\varepsilon_\mathrm{F}}\mathrm{d}\varepsilon\sum_{mm'm''m'''}\Delta_{i}^{mm'}G_{ij,\downarrow}^{m'm''}(\varepsilon)\Delta_{j}^{m''m'''}G_{ji,\uparrow}^{m'''m}(\varepsilon) \quad .
\end{equation}
\end{widetext}
Here, the positive (negative) sign applies if the spins $i$ and $j$ are aligned (anti-)parallel in the reference configuration, $G_{ij,\sigma}^{mm'}(\varepsilon)$ is the intersite Green's function, and $\Delta_{i}^{mm'}$ is the local exchange splitting on site $i$:
\begin{equation}\label{Eq:Delta}
    \Delta_{i}^{mm'} = H_{ii,\uparrow}^{mm'}-H_{ii,\downarrow}^{mm'} \quad ,
\end{equation}
with $H_{ij,\sigma}^{mm'}$ being the Hamiltonian in the tight-binding-like basis.

To obtain these quantities from our density functional theory (DFT) calculations, we follow the computational scheme used in Refs.~\onlinecite{Rudenko:2013,korotin:2015,logemann:2017}, and express the Kohn-Sham Hamiltonian in terms of site-centered maximally localized Wannier functions~\cite{MLWF}, $|w_{i\sigma}^m(\bf{R})\rangle$. Here, $i$ denotes only sites within the unit cell with lattice vector $\bf{R}$, where the Wannier function with orbital character $m$ and spin projection $\sigma$ is located. The Kohn-Sham Hamiltonian in the Wannier basis is then denoted as $H_{ij,\sigma}^{mm'}(\mathbf{R})$, where $\bf{R}$ is the lattice vector connecting the unit cells where the two Wannier functions are located. The Hamiltonian is spin-diagonal, since we are considering only collinear reference configurations and neglect spin-orbit coupling.

The Green's function matrix in reciprocal space is:
\begin{equation}\label{eq:green1}
    G_{\sigma}(\varepsilon,\mathbf{k}) = [\varepsilon - H_{\sigma}(\mathbf{k})]^{-1} \quad ,
\end{equation}
where $H_{\sigma}(\bf{k})$ is the reciprocal-space Hamiltonian matrix obtained from Fourier transformation of $H_{\sigma}(\bf{R})$.

The real space Green's function can then be obtained via integration over the Brillouin zone (BZ) 
\begin{equation}
    G_{ij,\sigma}^{mm'}(\varepsilon,\Delta\mathbf{R}) = \frac{1}{N_k} \sum_{\mathbf{k}} G_{ij,\sigma}^{mm'}(\varepsilon,\mathbf{k})e^{i\mathbf{k}\cdot \Delta\mathbf{R}} \quad ,
\end{equation}
where $\Delta\mathbf{R}$ is the lattice vector connecting the two unit cells where the corresponding Wannier functions are located, and $N_k$ is the number of $\bf{k}$-points in the Brillouin zone.

One advantage of this method is that it allows us to compute orbital resolved contributions to the exchange interactions. Due to the cubic symmetry, the exchange splitting matrix $\Delta_i$ is diagonal in our calculations. Thus, one can directly obtain the exchange interaction between orbital $m$ on site $i$ and orbital $m'$ on site $j$ as~\cite{korotin:2015}:
\begin{equation}\label{eq:per_ob}
    J_{ij}^{mm'} = \pm \frac{1}{2\pi} \mathrm{Im} \int_{-\infty}^{\varepsilon_\mathrm{F}}\dd\varepsilon \ \Delta_{i}^{mm}
    G_{ij,\downarrow}^{mm'}\Delta_{j}^{m'm'}G_{ji,\uparrow}^{m'm} \quad .
\end{equation}
The sum over all orbital contributions $m$ and $m'$ then gives the total exchange coupling between $i$ and $j$.
Furthermore, one can also analyze the energy dependence of the integrand in Eq.~\eqref{eq:J_ij} to identify the origin of different contributions to $J_{ij}$ .

The important difference between the two methods to calculate $J_{ij}$ is that Eq.~\eqref{eq:Jij-energy} is based on total energy differences for fully flipped spins, whereas Eq.~\eqref{eq:J_ij} corresponds to infinitesimal spin rotations away from the reference configuration. 
In the following, we will always refer to Eq.~\eqref{eq:Jij-energy} as the total energy method and to Eq.~\eqref{eq:J_ij} as the method of infinitesimal spin rotations (ISR).

If a material is well described by the Heisenberg Hamiltonian, both methods should give identical results and, furthermore, these results should be independent of the chosen magnetic reference state. In reality it is well known, however, that for many materials, in particular itinerant metallic magnets, the Heisenberg model is not necessarily a good approximation~\cite{Moriya:Book}, and thus results obtained by the two methods may differ, and furthermore can also depend on the chosen reference configuration. 
In this case, the exchange couplings obtained using the method of ISR,  Eq.~\eqref{eq:J_ij}, can still give a good description of transversal low-energy spin fluctuations around the reference state, whereas the total energy method might be more suitable to parameterize energy differences between different magnetic phases.

The Heisenberg Hamiltonian can be derived as a model describing low energy excitations of the Hubbard model in the insulating limit,~\cite{Fazekas:1999} and thus insulating magnetic oxides, such as SrMnO$_3$, are often thought of being well-described by the Heisenberg model.
However, previous work has reported pronounced non-Heisenberg behavior for SrMnO$_3$~\cite{Edstrom/Ederer:2018} and also for other manganites~\cite{Fedorova_et_al:2015}.

\subsection{Technical details}

 We perform spin polarized DFT+$U$ calculations \cite{DFTU} with the Vienna \textit{ab initio} simulation package (VASP) \cite{VASP1,VASP2,VASP3} using the projector augmented wave (PAW) method \cite{PAW1,PAW2}. Similar to previous studies of SrMnO$_{3}$~\cite{Marthinsen_et_al:2016,Edstrom/Ederer:2018}, we use the PBEsol~\cite{PBEsol} exchange-correlation functional, and a Coulomb repulsion\cite{PhysRevB.57.1505} $U_{\mathrm{eff}} = 3~\mathrm{eV}$ is added on the Mn $d$-electrons. 
 For comparison, we also perform some total energy calculations using the hybrid functional according to Heyd, Scuseria, and Ernzerhof (HSE)~\cite{Heyd/Scuseria/Ernzerhof:2003}.
 
 A plane wave energy cut-off of 680~eV and a $7\times 7 \times 7$ $\bf{k}$-point grid are used in combination with a doubled perovskite unit cell with lattice vectors $a(0,1,1)$, $a(1,0,1)$ and $a(1,1,0)$ to accommodate the G-type antiferromagnetic order. For the HSE calculations, an $8\times 8 \times 8$ $\bf{k}$-point grid was used together with a two-fold down-sampling of the $\bf{k}$-points and a reduction in the FFT grid used for the exact exchange, allowing for computational speed up. Comparison to calculations without such down-samplings indicate an error of well below 1\% for the total energy difference between the FM and G-AFM states. Identical crystal structures are used to compare ferromagnetic (FM) and antiferromagnetic (AFM) reference states, and the equilibrium lattice constant of the basic perovskite unit cell is $a_{0} = 3.79$ {\AA}, according to the PBEsol+$U$ calculations for the G-AFM state~\cite{Edstrom/Ederer:2018}.

The Wannier functions of SrMnO$_{3}$ are constructed from initial projections on Mn 3$d$ and O 2$p$ bands in the relevant energy range using the WANNIER90 code~\cite{wannier90}, resulting in 28 basis functions per spin channel for the doubled perovskite unit cell containing two Mn and six O atoms.
Thus, for each $\bf{R}$ and $\sigma$, the Hamiltonian is a $28 \times 28$ matrix.
For Fourier transforming quantities in the Wannier basis, we define a $\bf{k}$-points mesh which does not necessarily coincide with that used for the DFT calculations. In the AFM state, SrMnO$_3$ is an insulator, so $10 \times 10 \times 10$ $\bf{k}$-points are found to be sufficient. However, in the FM state, SrMnO$_3$ is a metal, where the states near the Fermi level greatly affect the convergence. Thus, in this case we use up to $30 \times 30 \times 30$ $\bf{k}$-points to achieve good convergence. 

According to our definition, Eq.~\eqref{eq:green1}, the Green's function has poles on the real axis. Thus, we substitute $\varepsilon$ by $\varepsilon+i\eta$, where $\eta>0$ is an infinitesimal smearing parameter. We then use Cauchy's theorem to evaluate the energy integral in Eq.~\eqref{eq:J_ij} in the limit $\eta \rightarrow 0+$ by integrating over a semi-circular contour in the upper half of the complex energy plane, starting at the Fermi level and ending below the bottom of the relevant bands, using 3000 sampling points. 
When analyzing the energy dependence of the integrand in Eq.~\eqref{eq:J_ij} (Fig. \ref{fig:f3}), we approximate the Green's function along the real axis using a small but finite value for $\eta>0$.

\section{Results}\label{sec:res}

\subsection{Exchange interactions under isotropic volume expansion}\label{sec:J1vol}

\begin{figure}
    \centering
    \includegraphics[width=1 \linewidth]{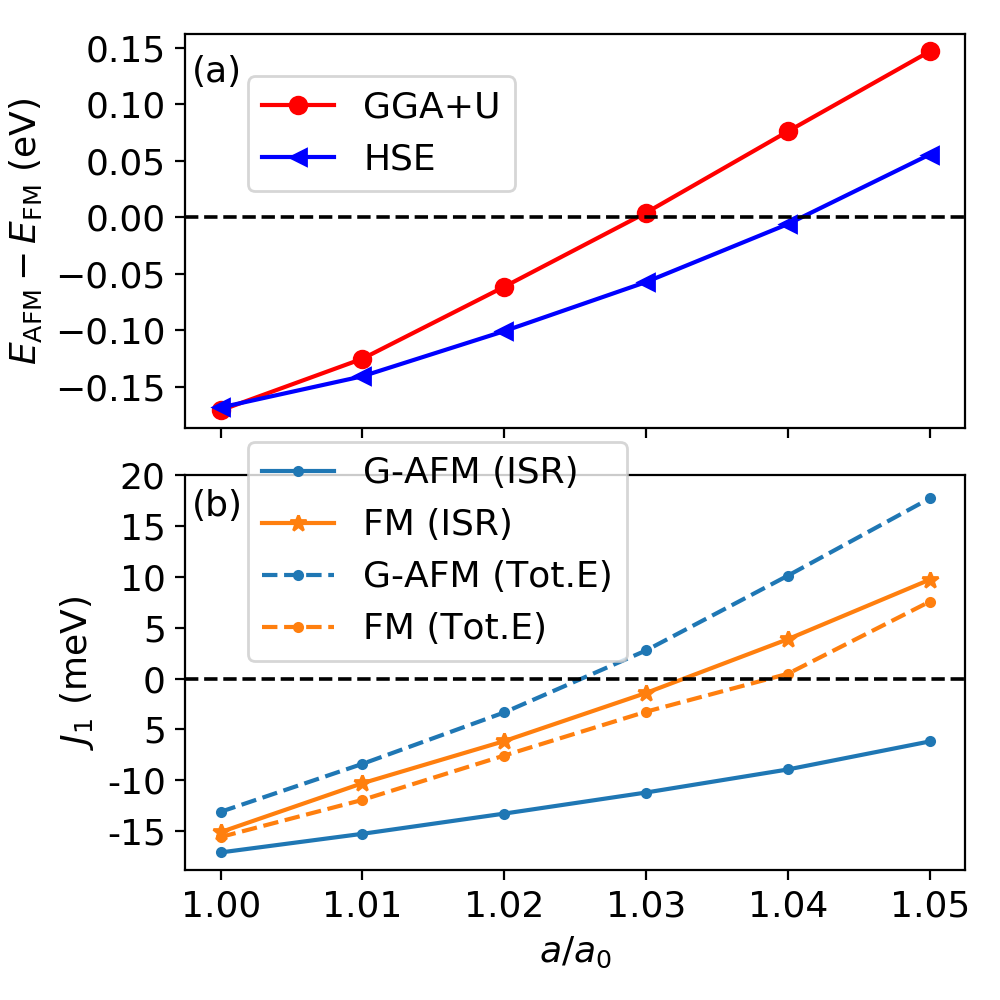}
    \caption{(a) Total energy difference (per formula unit) between G-AFM and FM states obtained from DFT calculations within GGA+$U$ (red) and HSE (blue), respectively, for cubic SrMnO$_3$ as a function of lattice constant. (b) Nearest neighbor exchange interaction $J_1$ as function of lattice constant for FM (orange) and G-AFM (blue) states obtained from the ISR approach (full lines) and from the total energy method (dashed lines) within GGA+$U$.}
    \label{fig:EJa}
\end{figure}

Fig. \ref{fig:EJa} (a) shows the total energy difference between the G-AFM and FM states as a function of lattice constant, i.e., under isotropic volume expansion, as obtained from our PBEsol+$U$ (denoted as GGA+$U$) and HSE calculations. In both cases the AFM state has lower energy than the FM state at the equilibrium lattice constant $a_0$, in agreement with experimental observations \cite{takeda1974}. However, the energy difference between FM and AFM states decreases with increasing volume, and for $a/a_{0} \geq 1.03$ the FM state becomes lower in energy. Thus, there is a phase transition from AFM to FM as the volume expands. 
A similar transition has also been obtained in previous DFT calculations of cubic SrMnO$_3$ under negative pressure~\cite{Chen_et_al:2014}.

The transition from AFM to FM order occurs for both PBEsol+$U$ and HSE, even though in the latter case it is shifted from $\sim 3$\,\% to $\sim 4$\,\% strain. This indicates that this magnetic transition is not an artefact of a potential gap underestimation in the case of the PBEsol+$U$ calculations. 
At the equilibrium lattice constant the energy difference between AFM and FM is essentially identical within PBEsol+$U$ and HSE. Note that  exactly this energy difference has been used to obtain reasonable $U$ values for SrMnO$_3$ and related systems in Ref.~\onlinecite{Hong_et_al:2012}. Thus, while the quantitative difference between PBEsol+$U$ and HSE increases with increasing lattice constant, both cases exhibit the same qualitative behavior, and in the following we therefore use only the computationally less-demanding PBEsol+$U$ approach. 

The trend observed for the total energy is mirrored (to large extent) in the exchange interactions.
Fig.~\ref{fig:EJa}(b) shows the nearest neighbor exchange interaction $J_1$ calculated for FM and G-AFM reference configurations using both the ISR (solid lines) and total energy (dashed lines) methods.
In all cases, $J_1$ is strongly negative at the equilibrium lattice constant, $J_1 \sim -15$\,meV, then decreases in strength with increasing volume, and eventually changes sign, indicating a transition from AFM to FM,  for strains around 3-4\,\%, except for the $J_1$ calculated for the G-AFM reference configuration using the ISR method, which remains negative over the whole range considered in our calculations.

This behavior in $J_1$ as function of interatomic distance is consistent with that previously observed for the in-plane nearest neighbor interaction when applying biaxial tensile strain~\cite{Edstrom/Ederer:2018}, and it appears reasonable to assume that the emerging FM coupling has the same origin in both cases. 
However, we point out that in the present case the system always remains cubic, i.e., all Mn-O-Mn bonds form an ideal 180$^\circ$ angle and all three $t_{2g}$-orbitals on the Mn atom remain degenerate. Thus, the sign change of $J_1$ from AFM to FM coupling under isotropic volume expansion is not straightforward to understand using conventional superexchange arguments.

For a system well described by the Heisenberg Hamiltonian, the four different ways of calculating $J_1$ shown in Fig.~\ref{fig:EJa}(b) should give the same result. Hence, the data shows increasing non-Heisenberg behavior with increasing volume, since the spread in the data becomes more pronounced for larger lattice constants. This spread is particularly large for the $J_1$ obtained with the ISR method. In this case, for large volume expansion, there is even a qualitative difference with different sign in $J_1$ depending on the magnetic reference state.
Since the exchange interactions obtained using ISR describe the curvature of the total energy for small deviations around the reference state, these results indicate local stability of the G-AFM configuration, i.e.,the presence of a local energy minimum. 

\begin{figure}
    \centering
    \includegraphics[width=1 \linewidth]{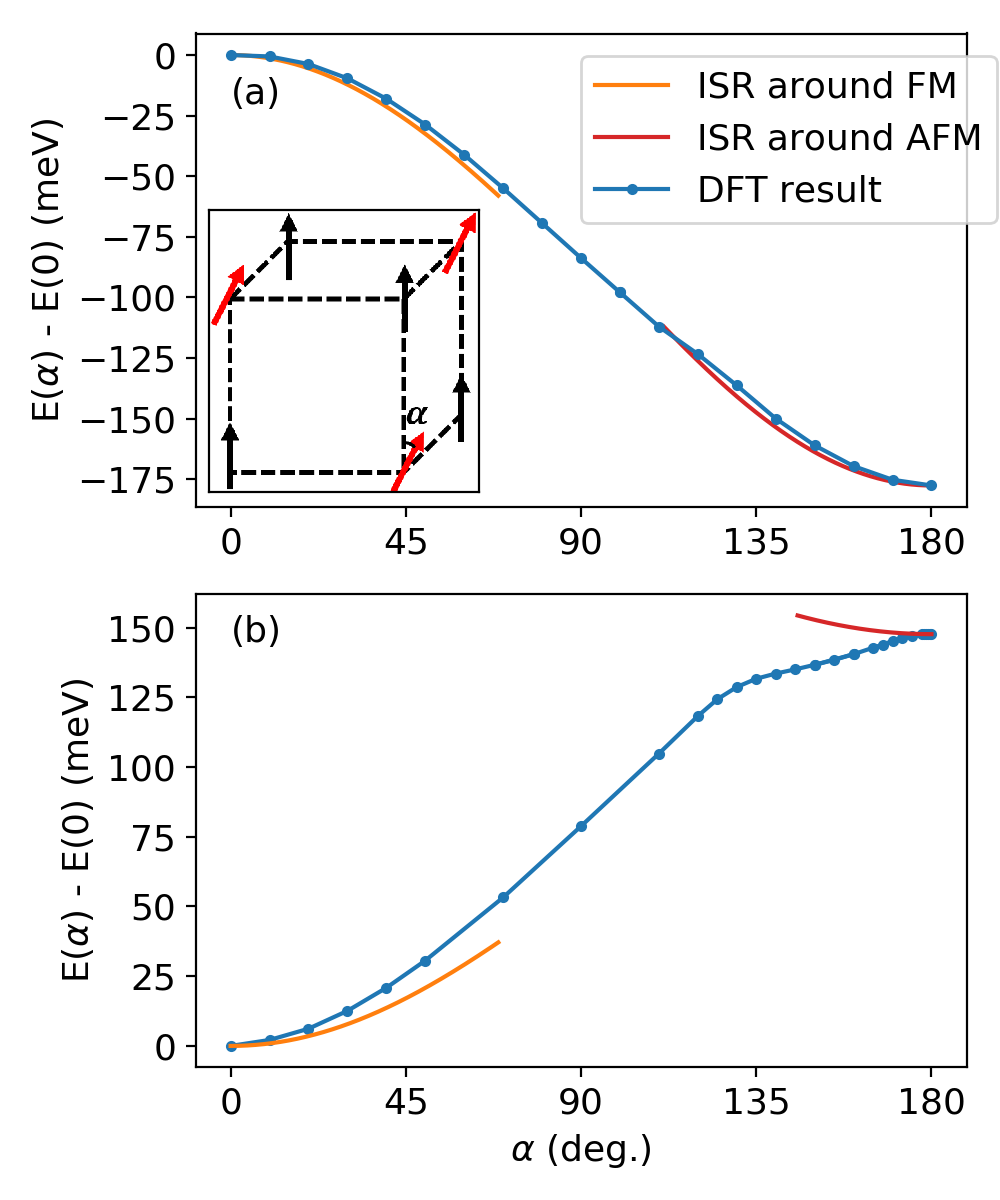}
    \caption{Total energy variation (per formula unit) as function of the spin rotation angle $\alpha$, as illustrated in the inset in (a). The FM state corresponds to $\alpha = 0^\circ$ and the G-AFM state to $\alpha=180^\circ$. Blue lines show DFT total energies, red and orange lines indicate the energy change around the FM and G-AFM reference configurations calculated using Eq.~\eqref{eq:ncl} with $J_1$ obtained from the ISR method for the corresponding reference state. Panel (a) corresponds to the equilibrium lattice constant $a/a_{0}=1$, whereas panel (b) corresponds to $a/a_{0}$=1.05.}
    \label{fig:ncl}
\end{figure}

To investigate this further, and to better understand the non-Heisenberg behavior, we now present results of non-collinear DFT calculations, where we continuously transform the system from the FM to the G-AFM configuration.
For this, we rotate one magnetic moment within the quasi-rhombohedral doubled perovskite unit cell by an angle $\alpha$ relative to the other one, as illustrated in the inset of Fig.~\ref{fig:ncl}(a).
Thus, $\alpha=0^\circ$ corresponds to the FM state and $\alpha=180^\circ$ corresponds to G-AFM. 

Based on the Heisenberg Hamiltonian, the energy per Mn atom as a function of rotation angle $\alpha$ is
\begin{equation}
\begin{split}\label{eq:ncl}
    E(\alpha) & = -\frac{1}{N}\sum_{i<j}J_{ij}\hat{m}_{i}\cdot\hat{m}_{j} \\
    & = -(6J_{1}+8J_{3})\cos\alpha+E_{0}
  \end{split}
\end{equation}
where $N$ is the number of Mn atoms, $J_n$ is the exchange coupling corresponding to the $n$-th nearest neighbors, $E_{0}$ is the sum over $\alpha$-independent terms, and up to fourth-nearest neighbors have been considered. 
As seen in Eq.~\eqref{eq:ncl}, the second- and fourth-nearest neighbor interactions do not contribute to the $\alpha$-dependence of the energy, and since $J_3$ has previously been found to be negligible~\cite{Edstrom/Ederer:2018}, it will not be considered in the following.

In Fig.~\ref{fig:ncl}(a), the DFT total energy calculated at the equilibrium lattice constant as a function of the spin rotation angle $\alpha$, varying from the FM to the G-AFM state, is shown (blue line). In addition, the expected energy variations around the FM and G-AFM states according to Eq.~\eqref{eq:ncl}, with $J_1$ obtained for the corresponding reference states using the ISR method, are indicated by the orange and red lines, respectively.
Here, the red and orange lines fit the total energy curve rather well. By fitting the DFT total energies to a cosine function, according to Eq.~\ref{eq:ncl}, we obtain $J_1^\mathrm{fit} = -15.2~\mathrm{meV}$.
This value can be compared to those calculated with the ISR method, of $J_1^\mathrm{FM} = -15.0~\mathrm{meV}$ and $J_1^\mathrm{AFM} = -17.1~\mathrm{meV}$, for the FM and AFM reference states, respectively. Thus, in spite of these moderate quantitative differences, the system is reasonably well described by the Heisenberg model at its equilibrium lattice constant.

This is not true any more under isotropic volume expansion. 
Fig.~\ref{fig:ncl}(b), shows analogous results as in (a), but calculated for an expanded volume with $a/a_{0}=1.05$. Clearly, the blue line does not follow the expected cosine behavior, in particular close to the AFM state at $180^\circ$, thus indicating strong non-Heisenberg behavior. 
Nevertheless, the orange line, corresponding to ISR around the FM state still agrees reasonably well with the total energy calculations for small deviations around $\alpha=0$. This is less obvious for the red line, corresponding to ISR around the G-AFM state. 
We note that the accuracy of the noncollinear constrained moment method does not allow to investigate arbitrary small rotation angles.
From the negative $J_1$ obtained using the ISR method, one would expect a local energy minimum for $E(\alpha)$ around $\alpha = 180^\circ$. While it is not clear that such a minimum appears in the total energy curve (blue line) in Fig.~\ref{fig:ncl}(b), the particularly strong deviation from Heisenberg behavior around the AFM state at $a/a_{0}=1.05$ explains the markedly different behavior in the calculated $J_1$ around this point. 

We point out that in the AFM configuration only the Mn atoms exhibit magnetic moments, whereas in the FM case also the O atoms carry small magnetic moments.
Since the O moments vanish in the AFM case, their nature is intrinsically non-Heisenberg.
Nevertheless, this raises the question of whether the apparent non-Heisenberg behavior of the Mn spins can be reduced by considering the spin-polarization of the O atoms in an appropriate way.
Several authors have discussed spin polarization on oxygen atoms as a potential source for non-Heisenberg behavior in transition metal oxides~\cite{doi:10.1143/JPSJ.78.054710,logemann:2017,PhysRevMaterials.2.073001,PhysRevB.97.184404}. 
Considering a spin polarization energy of the O atoms, as discussed, e.g., in Ref.~\onlinecite{PhysRevB.97.184404}, will affect the individual energies entering Eq.~\eqref{eq:Jij-energy}, but will contribute equally for the FM and AFM reference states (assuming that the size of O magnetic moment depends only on the neighboring Mn spins) and thus can not explain the observed configuration dependence. 
In Refs.~\onlinecite{logemann:2017,PhysRevMaterials.2.073001}, an extended Heisenberg model including couplings between Mn and O spins has been considered, and ``effective'' couplings between transition metal spins incorporating the effect of the O moments have been defined. 
However, such a model, applicable only for the FM case, can not explain the configurational dependence observed in the bare Mn-Mn exchange interactions calculated with the ISR method.
Furthermore, we have also verified that considering the effective Mn-Mn nearest neighbor interactions (incorporating the closest neighbor Mn-O couplings) obtained within the ISR method for the FM case does not improve the agreement with the $J_{ij}$'s obtained for the AFM reference configuration.

\subsection{Orbital- and energy-resolved contributions to the exchange interaction}
\label{sec:orbres}

\begin{figure}
    \centering
    \includegraphics[width=1 \linewidth]{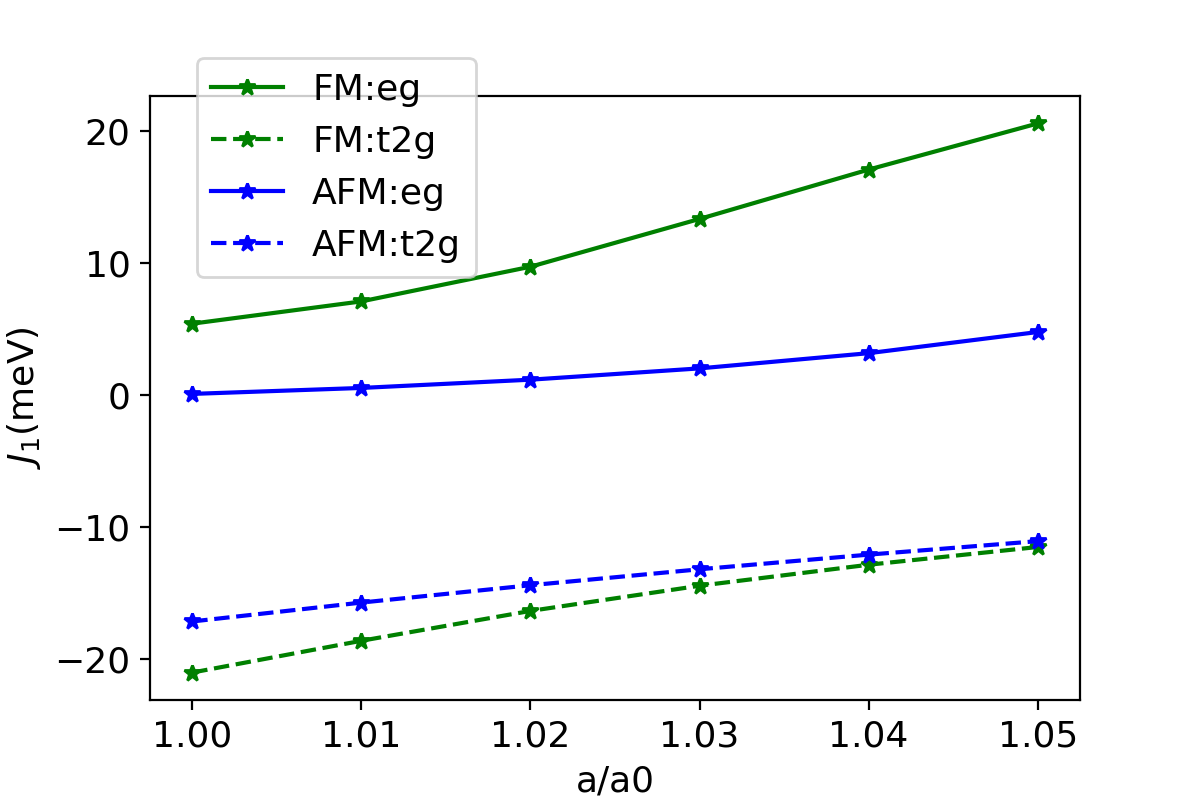}
    \caption{Orbital-resolved contributions to the nearest neighbor exchange interaction $J_1$ as function of lattice constant. Green lines show the results for FM order, while the blue lines show AFM order. Full (dashed) lines correspond to the $e_g$ ($t_{2g}$) contribution.}
    \label{fig:Job}
\end{figure}

To understand the origin of the sign change of the magnetic nearest neighbor interaction under isotropic volume expansion and of the strong non-Heisenberg behavior, we apply Eq.~\eqref{eq:per_ob} to obtain orbital-resolved contributions to $J_{1}$.
Since we are considering interactions between Mn $d$-electrons in an octahedral crystal field, these can be decomposed into contributions of $e_g$ or $t_{2g}$ character, while mixed $e_{g}$-$t_{2g}$ contributions are zero by symmetry. 
Fig.~\ref{fig:Job} shows the total $e_{g}$ and $t_{2g}$ contributions to $J_1$ under isotropic volume expansion as functions of lattice constant. 
The data shows that in all cases the $t_{2g}$ contribution is negative while the $e_g$ contribution is positive.
For the G-AFM reference state, the $e_g$ contribution to the exchange coupling is negligible at the equilibrium lattice constant and remains small over the whole range up to $a/a_0=1.05$. 
In contrast, for the FM reference state, the $e_g$ contribution is non-negligible already at $a_0$ and increases strongly with increasing volume, while simultaneously the $t_{2g}$ contribution decreases in strength. Thus, the sign change of $J_1$ in the FM state is due to the strong positive $e_g$ contribution dominating at large volumes.
The large difference of the $e_g$ contribution for the FM and AFM reference states is also the reason for most of the non-Heisenberg behavior of $J_1$, although there is also a smaller discrepancy in the magnitude of the $t_{2g}$ contributions near the equilibrium volume.
The results in Fig.~\ref{fig:Job} thus show that the $e_g$ states are crucial for both the non-Heisenberg behavior and the transition from AFM to the FM state with increasing volume in SrMnO$_3$.

\begin{figure}
    \centering
    \includegraphics[width=1 \linewidth]{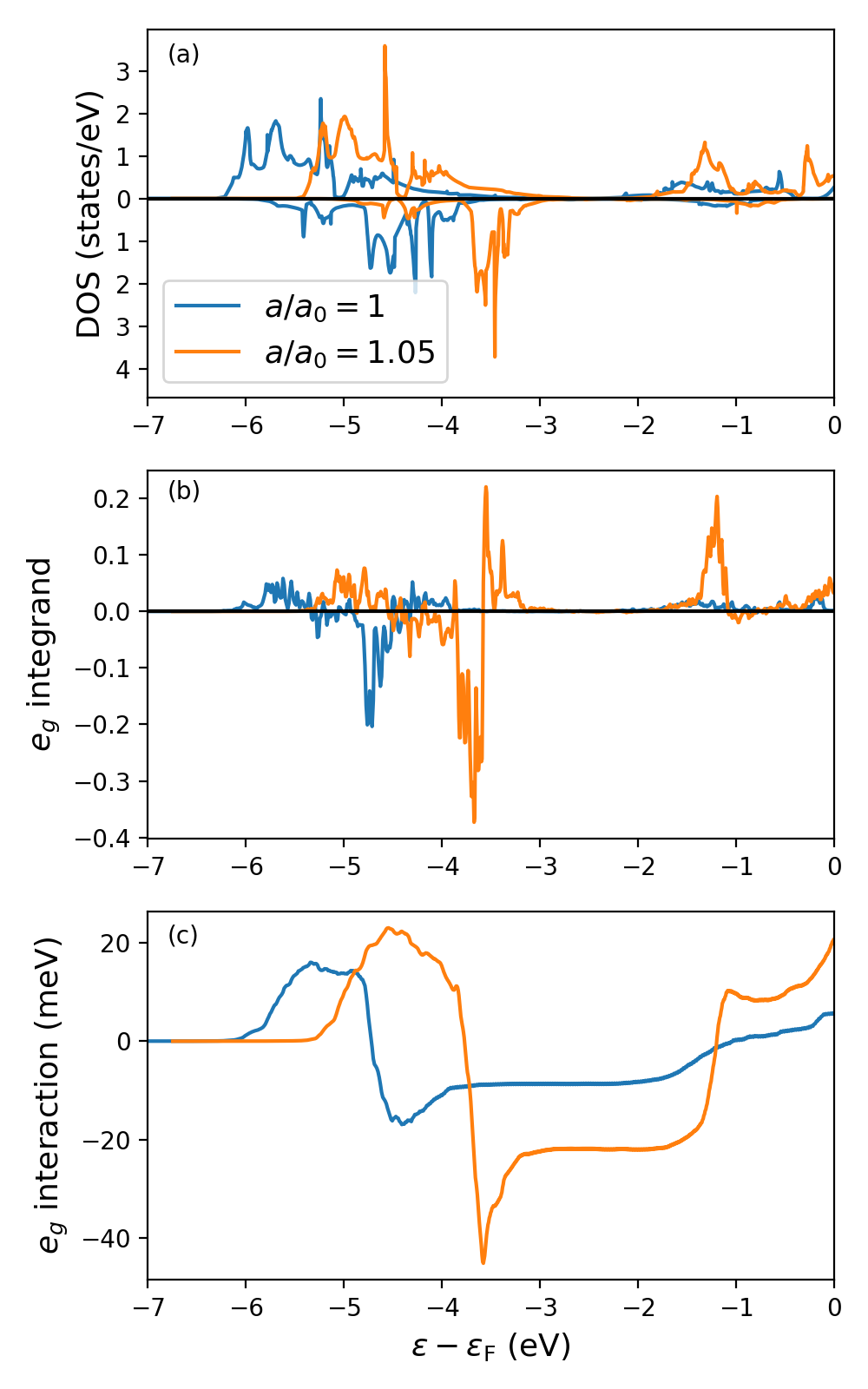}
    \caption{(a) Mn $e_{g}$ contribution to the density of states (DOS), with spin up on the positive axis and spin down on the negative axis. (b) Energy-resolved $e_{g}$ contribution to the integrand of $J_{1}$ according to Eq.~\eqref{eq:eg-integrand}. (c) $e_{g}$ contribution to $J_{1}$, integrated up to an energy $\varepsilon \leq \varepsilon_\text{F}$ (Eq.~\eqref{eq:eg-int}). All data in (a)-(c) corresponds to the FM state calculated for lattice constants $a/a_{0}=1$ (blue) and $a/a_{0}=1.05$ (orange).}
    \label{fig:f3}
\end{figure}

To further elucidate the reason behind the strong FM character of the $e_{g}$ contribution to the nearest neighbor exchange interaction, Fig.~\ref{fig:f3}(b) shows the energy-resolved $e_g$ contribution to the integrand of Eq.~\eqref{eq:per_ob}, i.e.:
\begin{equation}
\label{eq:eg-integrand}
    j^{e_g}_{ij}(\varepsilon) = \pm \frac{1}{2\pi}\sum_{mm'}\mathrm{Im}[\Delta_{i}^{mm}G_{ij,\downarrow}^{mm'}(\varepsilon)\Delta_{j}^{m'm'}G_{ji,\uparrow}^{m'm}(\varepsilon)] \quad ,
\end{equation}
where $m$ and $m'$ run through the $e_{g}$ orbitals $|3z^{2}-r^{2}\rangle$ and $|x^{2}-y^{2}\rangle$, and $ij$ corresponds to nearest neighors.
For comparison, Fig.~\ref{fig:f3}(a) contains the $e_g$-projected density of states (DOS) below the Fermi level. 
In addition, Fig.~\ref{fig:f3}(c) shows the $e_g$ contribution to $J_1$, integrated up to a certain energy $\varepsilon \leq \varepsilon_\text{F}$:
\begin{equation}
\label{eq:eg-int}
    J_{ij}^{e_g}(\varepsilon) = \int_{-\infty}^{\varepsilon} d\varepsilon' j^{e_g}_{ij}(\varepsilon') \quad ,
\end{equation}
Thus, $J_{ij}^{e_g} = J_{ij}^{e_g}(\varepsilon = \varepsilon_\text{F})$.
All sub-figures correspond to the FM reference state.

It can be seen that the energies with strong contributions to the exchange interaction approximately match the energies where the $e_g$ DOS is large (even though there is no one-to-one correspondence between these two quantities). 
There are strong contributions to both DOS and the $e_g$ integrand 
at energies between approximately $-$6\,eV and $-3$\,eV.  These result from strong hybridization between the Mn $e_{g}$ orbitals with the O $p$ bands, which are located in that energy region. Additionally, there are $e_{g}$ contributions between approximately $-$2\,eV and the Fermi level, which become more pronounced for $a/a_{0}=1.05$. These are the contributions that are responsible for the positive sign of the $e_g$ contributions to $J_1$, as can be seen from the integrated quantity in Fig.~\ref{fig:f3}(c), which is still negative at $\varepsilon=-2$\,eV, but then becomes positive in the energy range between $-2$\,eV and the Fermi level.
The increase of the positive $e_g$ contributions in this energy range for $a/a_0=1.05$ is related to a lowering of the $e_g$-dominated bands for increasing lattice constant.

\subsection{Further neighbor interactions}\label{sec:furtherneigh}

\begin{figure}
    \centering
    \includegraphics[width=1 \linewidth]{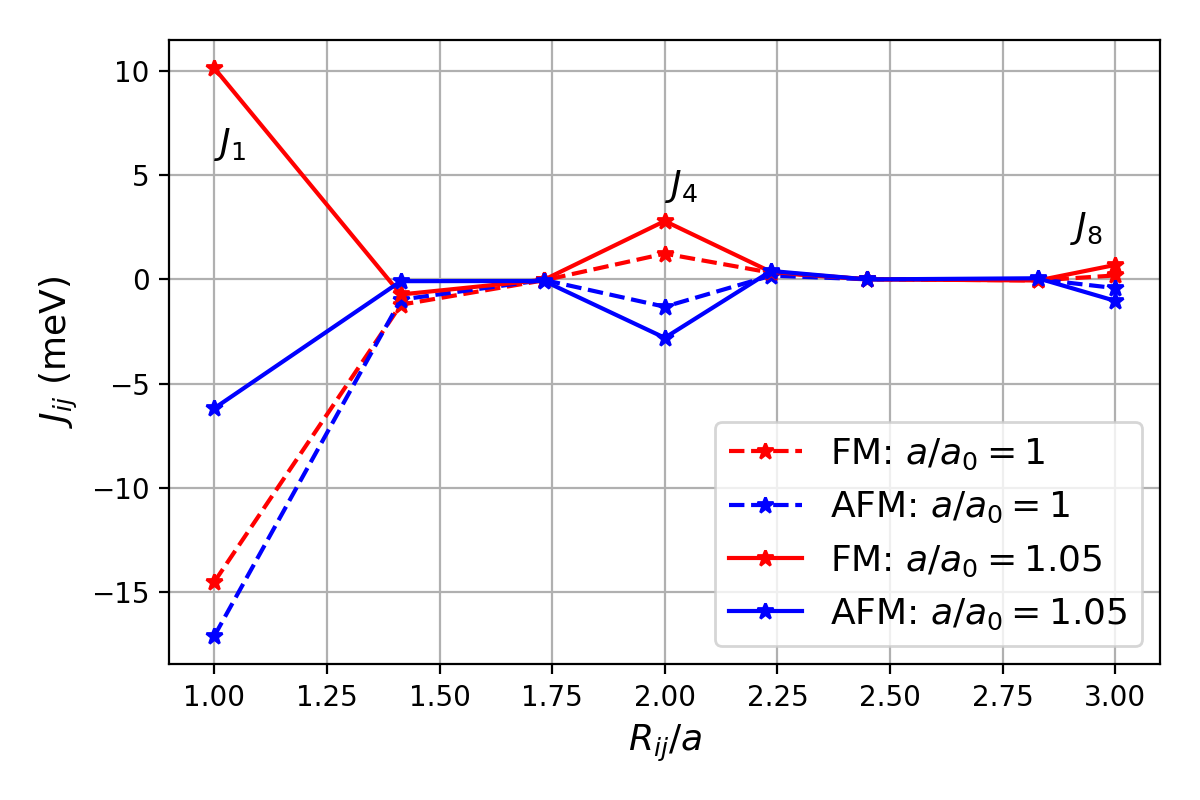}
    \caption{Further nearest neighbor exchange interaction as a function of distance obtained using the ISR method. The figure shows exchange interaction for FM (red) and AFM (blue) order at lattice constants $a/a_{0}=1$ (dashed lines) and $a/a_{0}=1.05$ (solid lines).}
    \label{fig:JR}
\end{figure}

Having analyzed the origin for the sign change of the nearest neighbor coupling, we now present results for further neighbor interactions.
Previously, total energy calculations have been used to calculate up to the third nearest neighbor exchange interactions (with distance $\sqrt{3}a$) in SrMnO$_3$~\cite{Edstrom/Ederer:2018}. 
The second and third neighbor interactions were found to be more than one order of magnitude smaller than $J_1$, while further neighbor interactions were assumed to be negligible.
An advantage of the ISR method is, that it allows to calculate long distance exchange interactions without the need for prohibitively large supercells. Fig.~\ref{fig:JR} shows Mn-Mn exchange interactions with interatomic distances up to $3a$.\footnote{Note that there are two distinct types of 8th nearest neighbors with distance $3a$ in the cubic lattice, corresponding to lattice vectors of type $(3,0,0)a$ and $(2,1,1)a$, respectively. In this work, we only consider the first type.}

It can be seen that the fourth and eighth neighbor exchange interactions $J_4$ and $J_8$ 
are notably larger than the second and third neighbor exchange interactions, as well as other calculated further neighbor interactions. $J_4$ and $J_8$ correspond to interactions along the cubic crystallographic axes, with interatomic distances of $2a$ and $3a$, respectively. These results indicate that it might be important to consider at least up to fourth neighbor exchange interactions in studies of magnetism in SrMnO$_{3}$. 
However, it is also apparent that both $J_4$ and $J_8$ exhibit pronounced non-Heisenberg behavior with different signs obtained for the FM and AFM reference states.

\begin{figure}
    \centering
    \includegraphics[width=1 \linewidth]{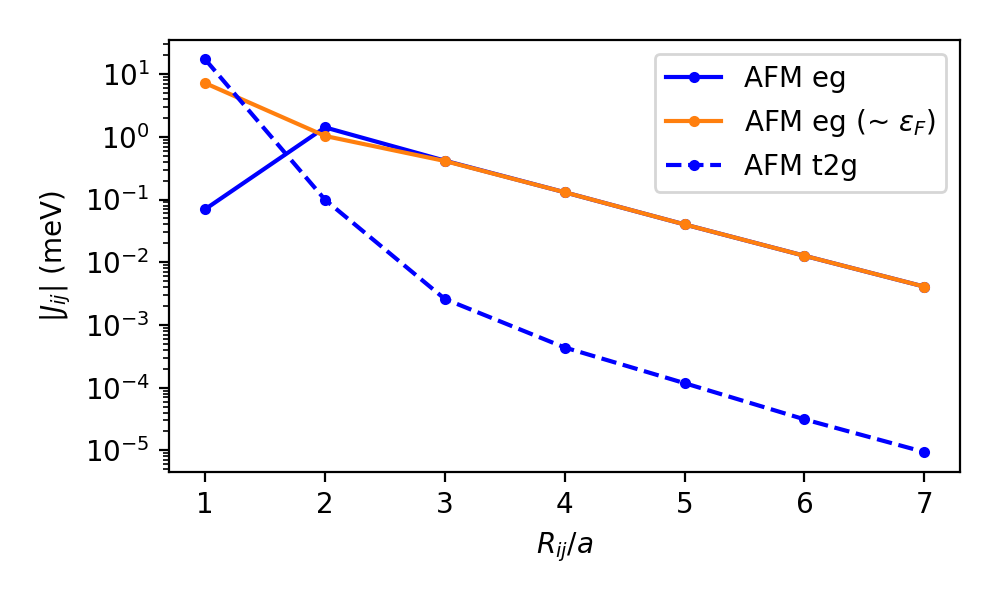}
    \caption{Absolute magnitude of the orbital resolved contributions to the exchange interactions along the [100] direction as function of interatomic distance $R_{ij}$, plotted on a logarithmic scale. The data is calculated for the AFM state at the equilibrium volume. The $e_g$ and $t_{2g}$ contributions correspond to the solid and dashed blue lines, respectively. The orange line shows the $e_g$ contribution obtained by evaluating the integral in Eq.~\eqref{eq:eg-int} only over an interval between approximately $-2$\,eV and the Fermi energy.}
    \label{fig:JR_log}
\end{figure}

Fig.~\ref{fig:JR_log} shows the magnitude of the orbital resolved contributions to the further neighbor exchange interactions along the cubic high symmetry [100]-direction on a logarithmic scale (calculated using the ISR method for the AFM state at the equilibrium volume). The linear decrease observed for large interatomic distances indicates an exponential decay of $|J_{ij}|$. The data in Fig.~\ref{fig:JR_log} also shows that the $e_g$ contribution (solid blue line) is several orders of magnitude larger than the $t_{2g}$ contribution (dashed blue line) for large distances. The orange line shows the $e_g$ contribution obtained when the integral in Eq.~\eqref{eq:eg-int} is evaluated only over a reduced energy interval between approximately $-2$\,eV and the Fermi level. For large interatomic distances, this data falls right on top of the one for the total $e_{g}$ contribution (integrated over all energies up to the Fermi energy). Hence, the occupied $e_g$ states near the Fermi energy are entirely responsible for the long-range behavior of the $e_g$ contribution, and hence also for the total exchange interaction.

\section{Summary and Conclusions}\label{sec:concl}

We have calculated magnetic exchange interactions in SrMnO$_3$, as function of lattice constant under isotropic volume expansion, using a method which evaluates energy variations due to infinitesimal spin rotations, and compared these results to calculations of total energy differences for collinear spin configurations. 
Previous work~\cite{Edstrom/Ederer:2018} showed that the in-plane nearest neighbor magnetic exchange interaction of \smo changes sign with increasing interatomic distance due to epitaxial strain.
This sign change is not easily understood from standard superexchange theory.
Here, we find that the same transition from AFM to FM coupling appears also under isotropic volume expansion. 

By analyzing the orbital-resolved and energy-dependent contributions to the nearest neighbor interaction $J_1$, we find that the positive sign of $J_1$ (favoring FM order) obtained for increasing interatomic distances stems from partial occupation of the Mn $e_{g}$ states near the Fermi level.
These Mn $e_g$ states are lowered in energy by the volume expansion, which enhances the hybridization between these states with the O $p$ bands, making a standard superexchange model less applicable.
Furthermore, the partial occupation of the $e_g$ states (at least in the FM case), in spite of the formal Mn$^{4+}$ valence configuration, can enable a double-exchange-like coupling mechanism, where itinerant $e_g$ electrons mediate a net FM coupling between more localized $t_{2g}$ spins~\cite{Millis/Littlewood/Shraiman:1995,Dagotto/Hotta/Moreo:2001}. 
Our results also indicate that the enhancement of the observed non-Heisenberg behavior under volume expansion originates mainly from these $e_{g}$ contributions.
This is similar to what has been reported for the metallic itinerant ferromagnet bcc Fe~\cite{Kvashnin2016}, even though this is a material rather different from SrMnO$_3$. 
As shown in Appendix \ref{AppA}, the same effect can also be observed for the strongly ionic antiferromagnetic insulator KMnF$_3$.

Our calculations of long range interactions in \smo reveal that the fourth and eighth neighbor interactions are significantly larger than the second or third neighbor interactions. Moreover, further analysis of the long range interactions show that while both $t_{2g}$ and $e_{g}$ contributions to the exchange interactions decay exponentially with interatomic distance, the long range behavior is dominated by the $e_{g}$ contributions stemming from the energy interval immediately below the Fermi level. 

To conclude, our results provide important insights into the mechanisms of magnetic exchange in SrMnO$_3$, which might also prove useful in future understanding of magnetism in other magnetic transition metal-oxides.
Our results show that pronounced non-Heisenberg behavior can occur for a magnetic insulator such a SrMnO$_3$, similar to what has been found for other manganites~\cite{Fedorova_et_al:2015}. This demonstrates the importance of choosing appropriate methods when mapping first principles electronic structure calculations on atomistic spin Hamiltonians. 

In Ref.~\onlinecite{PhysRevB.97.184404}, it was discussed how inclusion of spin dependence in the exchange-correlation functional used in DFT+$U$ calculations can affect the configurational dependence of the calculated magnetic exchange interactions. It was argued that excluding the spin dependence in the exchange-correlation functional, while inducing magnetic order only via the Coulomb repulsion $U$, leads to a more Heisenberg-like behaviour. 
For future studies, it could be of interest to investigate such effects also in SrMnO$_3$.
Furthermore, it might also be of interest to consider the possible effect of higher order exchange interactions~\cite{Fedorova_et_al:2015}. Since \smo is also known to become ferroelectric under strain, with an off-centrosymmetric structural distortion, it would also be of interest to use a generalized version of the method used here to consider the effect of spin-orbit coupling and evaluate anti-symmetric Dzyaloshinskii-Moriya exchange interactions.

\begin{acknowledgments}
This work was supported by the Swiss National Science Foundation (project code 200021E-162297) and the German Science Foundation under the priority program SPP 1599 (``Ferroic Cooling'').
Computational work was performed on resources provided by the Swiss National Supercomputing Centre (CSCS) and the ETH Z\"urich.
\end{acknowledgments}

\appendix
\section{KMnF$_{3}$}\label{AppA}

To test whether a more ionic and strongly insulating solid shows a behavior in better accordance with the Heisenberg Hamiltonian, we also study the magnetic exchange interactions of KMnF$_3$, which is an AFM insulator with a large band gap~\cite{Scatturin_et_al:1961,Wang_et_al:2019}
Interestingly, we find clear deviations from Heisenberg behavior also in this compound.
Furthermore, the non-Heisenberg behavior of the nearest neighbor exchange interaction originates entirely from the $e_{g}$-interactions, while the $t_{2g}$ interactions are configuration independent.

We calculate the electronic structure of KMnF$_3$ with the same methods as used for SrMnO$_3$, i.e., using the PBEsol exchange correlation functional implemented within the VASP code. The plane wave energy cut-off is set to 750\,eV and $7\times 7\times 7$ $\bf{k}$-points are used. The lattice parameter of cubic G-type anitiferromagnetic KMnF$_{3}$ is calculated to be $a_{0} = 4.18$\,\AA, which agrees well with the experimental value of 4.19\,\AA \cite{Okazaki:1961}. 
As for SrMnO$_{3}$, we investigate the nearest neighbor Mn-Mn exchange interaction as a function of the cubic lattice constant, using both the ISR method and the total energy method. 
For the total energy method, we use a $2\times2\times2$ supercell containing 8 Mn atoms. For the ISR method, a smaller cell with lattice vectors $\frac{a}{2}(0,1,1)$, $\frac{a}{2}(1,0,1)$ and $\frac{a}{2}(1,1,0)$, containing two Mn atoms, is used, which allows for the G-type AFM order. The Wannier functions are constructed from initial projections on Mn-centered $d$ and F-centered $p$ orbitals.

\begin{figure}
    \centering
    \includegraphics[width = 1\linewidth]{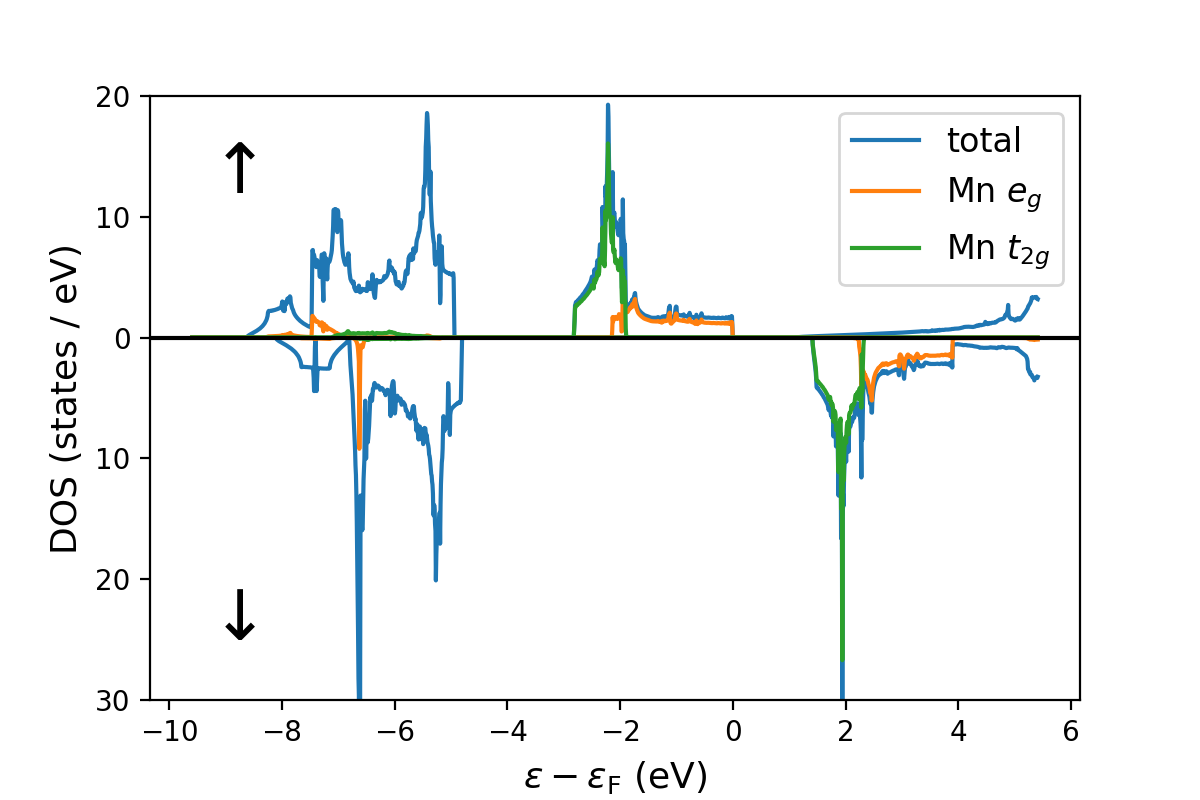}
    \caption{Total and orbital resolved DOS of FM KMnF$_{3}$ obtained from DFT. Spin up and spin down are shown along the positive and negative axes, respectively.}
    \label{fig:dos_KMF}
\end{figure}

Fig.~\ref{fig:dos_KMF} shows the DOS for FM KMnF$_{3}$ obtained from our DFT calculation. There is a strong splitting between spin up and spin down Mn $d$ orbitals present already without using an additional Coulomb repulsion $U$ in the DFT calculation. 
There are five unpaired $d$ electrons on the Mn cation forming a high spin state, three electrons occupying $t_{2g}$ orbitals and two occupying the $e_{g}$ orbitals. 
Due to the more ionic and strongly insulating character of KMnF$_{3}$, it can be expected to be a better Heisenberg magnet than SrMnO$_3$, since the Heisenberg model can be derived from the electronic Hubbard model in the insulating limit ~\cite{Fazekas:1999}, where the hopping amplitude $t$ is small compared to the large Coulomb repulsion, $t/U \rightarrow 0$.

\begin{figure}
\centering
\includegraphics[width=1\linewidth]{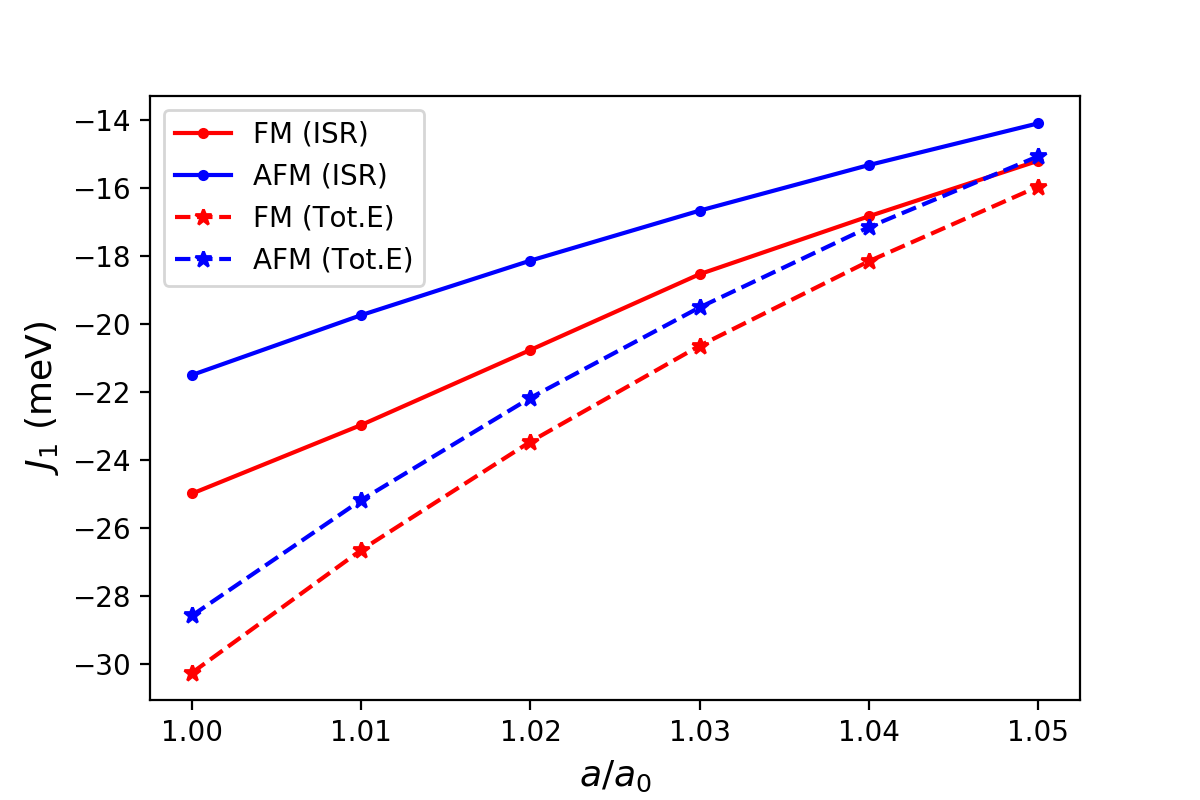}
\caption{Nearest neighbor exchange interaction in KMnF$_{3}$ as a function of lattice constant, calculated using the total energy method (dashed lines) and ISR method (solid lines) for a FM (red lines) and AFM (blue lines) reference configuration.}
\label{fig:J_KMF}
\end{figure}

Fig.~\ref{fig:J_KMF} shows the nearest neighbor exchange interaction $J_1$ as function of lattice constant obtained from the different methods for evaluating the exchange interaction, using both FM and G-AFM reference states. In each case, we obtain a negative $J_1$, favoring the AFM state. As the volume expands, the strength of the exchange interaction decreases, which can be understood from the increasing interatomic distances. In contrast to SrMnO$_{3}$, there is no sign change in $J_{1}$ within the considered range of lattice parameters. 

Both methods for calculating the exchange interactions exhibit differences between the FM (red lines) and AFM (blue lines) magnetic reference states, indicating deviations from the behavior of an ideal Heisenberg system also for KMnF$_{3}$. 
The difference between the coupling obtained from FM and AFM states is larger for the ISR method.
In addition, the corresponding coupling constants are weaker than the ones obtained using the total energy method.
This could be due to further neighbor couplings, which effectively contribute to the nearest neighbor interaction in the total energy method, due to the small size of the supercell.
This could also explain why the results of the two methods become more similar for larger volume, since the relative strength of these further neighbor couplings can be expected to decrease more rapidly with increasing interatomic distance.

\begin{figure}
\centering
\includegraphics[width=1\linewidth]{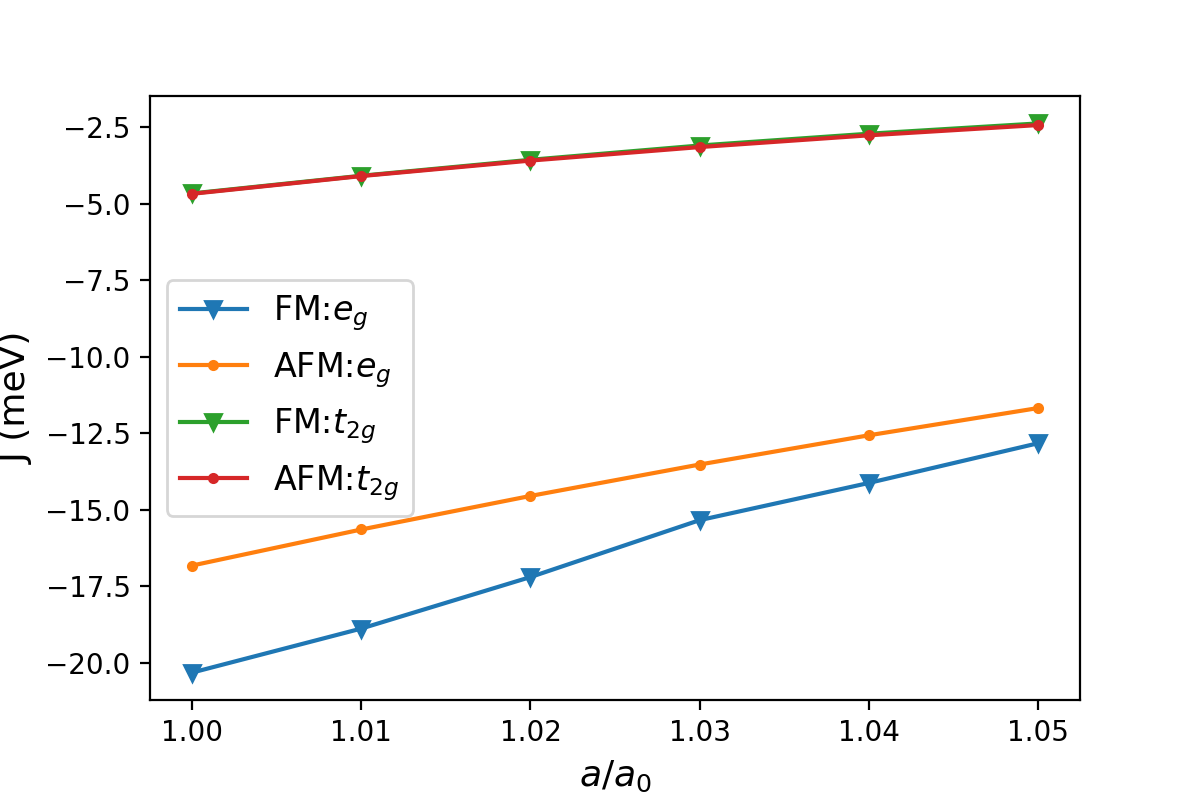}
\caption{Orbital resolved exchange interactions in KMnF$_{3}$ as a function of lattice constant. $t_{2g}-t_{2g}$ contribution for FM (green line) and AFM (red line) overlap, while $e_{g}-e_{g}$ contribution for FM (blue line) and AFM (orange line) show clear difference.}
\label{fig:Job_KMF}
\end{figure}

Finally, Fig.~\ref{fig:Job_KMF} compares the $e_{g}$ and $t_{2g}$ contribution for the FM and AFM states in the ISR method. A notable feature is the good overlap between the $t_{2g}$ contributions for both reference states, indicating excellent Heisenberg behavior. Interestingly, all the non-Heisenberg behavior stems from the $e_{g}$ interaction, similar to what has been observed for \smo in the main text and also for bcc Fe in a previous study~\cite{Kvashnin2016}.

\bibliography{bibfile}

\end{document}